\documentclass[runningheads]{llncs}
\usepackage{subfloat}
\usepackage{graphicx}
\usepackage{floatrow}
\usepackage{multirow}
\usepackage{authblk}
\usepackage{amsmath}
\usepackage{setspace}
\usepackage[super]{nth}
\usepackage[colorlinks=False,urlcolor=purple]{hyperref} 
\makeatletter
\newcommand{\printfnsymbol}[1]{%
  \textsuperscript{\@fnsymbol{#1}}%
}
\makeatother
\newfloatcommand{capbtabbox}{table}[][\FBwidth]

\begin{document}
%
\title{NuClick: From Clicks in the Nuclei to Nuclear Boundaries}
\titlerunning{NuClick}
\author[2]{Mostafa Jahanifar\thanks{These authors contributed equally to this work.}}
\author[1,3]{Navid Alemi Koohbanani\printfnsymbol{1}}
\author[1,3]{Nasir Rajpoot}

\affil[1]{Department of Computer Science, University of Warwick, Coventry}
\affil[2]{Department of Research \& Development, NRP Co., Tehran, Iran}
\affil[3]{Alan Turing Institute, London, UK}

\authorrunning{Jahanifar et al.}

\institute{}

%
%

\toctitle{Lecture Notes in Computer Science}
\tocauthor{Authors' Instructions}
\maketitle
\vspace{-0.5cm}
\begin{abstract}
Best performing nuclear segmentation methods are based on deep learning algorithms that require a large amount of annotated data. However, collecting annotations for nuclear segmentation is a very labor-intensive and time-consuming task.
Thereby, providing a tool that can facilitate and speed up this procedure is very demanding. Here we propose a simple yet efficient framework based on convolutional neural networks, named NuClick, which can precisely segment nuclei boundaries by accepting a single point position (or click) inside each nucleus.
Based on the clicked positions, inclusion and exclusion maps are generated which comprise 2D Gaussian distributions centered on those positions. These maps serve as guiding signals for the network as they are concatenated to the input image.  The inclusion map focuses on the desired nucleus while the exclusion map indicates neighboring nuclei and improve the results of segmentation in scenes with nuclei clutter. 
The NuClick not only facilitates collecting more annotation from unseen data but also leads to superior segmentation output for deep models. It is also worth mentioning that an instance segmentation model trained on NuClick generated labels was able to rank \nth{1} in LYON19 challenge.

\end{abstract}
%
%
%

\vspace{-0.9cm}
\section{Introduction}
\vspace{-.3cm}
Appearance and shape characteristics of nuclei in histology images can be considered as an important marker for the diagnosis of cancer and predicting patient outcome \cite{lu2018nuclear}. To quantify these features, one should first determine the boundaries of nuclei, which requires lots of time and effort to achieve manually.
To this end, automatic segmentation methods are playing an important role to facilitate tاhis task.

Since the emerge of deep learning (DL) methods and their superior performance over classical methods (feature-based), the need for annotated data has ramped up significantly.  Data-dependency nature of DL methods still imposes a huge burden on the human for providing annotated data.
Despite the labor intensity of annotating nuclei within histology images, several datasets have been provided for training deep networks \cite{kumar2017dataset,vu2019methods,koohbanani2019nuclear}. The question here is how we can use available annotated datasets to ease extending the knowledge and reducing the human effort when creating a new data set on another cancer/tissue type? In the computer vision domain, several methods have been employed to speed up the procedure of collecting annotations for natural images by accepting a few points from the annotator \cite{maninis2018deep}. One of the most efficient models is DEXTR \cite{maninis2018deep} which takes extreme points (the leftmost, rightmost, top and bottom pixels) of the object as the input to extract mask of the desired object. 

All these approaches require several points from the user to be clicked on the boundary of an object in order to delineate boundaries or draw a bounding box. For nuclear segmentation, providing several points on the boundaries of nuclei is still a high burden, since annotator should firstly find the boundary of a nucleus in high magnification and then select several points on it. Moreover, nuclei are small objects, and their number may exceed 400 in a patch size of 500$\times$500 (for example, when there is a dense cluster of lymphocytes), which make this task more arduous. To the best of our knowledge, there is no similar approach based on DL models for interactive nuclei segmentation in histology images. Some works like \cite{yang2006nuclei} used the marker-controlled watershed algorithm to segment nuclei from marked points which fails in complex histology images.

\begin{figure}[t]
    \centering
    \includegraphics[width=\textwidth]{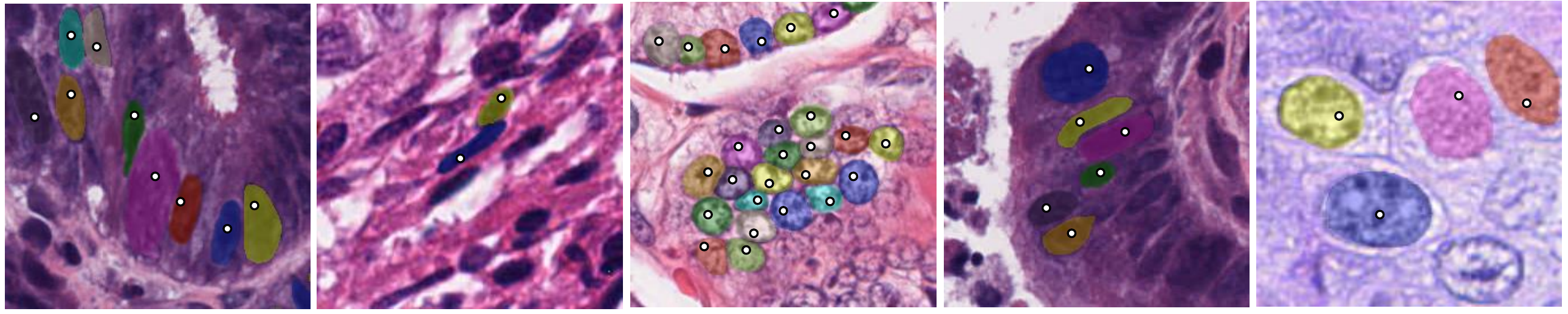}
    \caption{Example outputs of NuClick: Annotator click inside the nucleus and the mask will be generated by NuClick model.\vspace{-.5cm}}
    \label{fig:motiv}
\end{figure}

Here, we propose a simple yet effective method for collecting nuclear annotation by asking a user to provide only one point inside the nucleus (examples are depicted in Fig. \ref{fig:motiv}). Clicking one point inside an object is not a demanding task and can be done in low resolution by a non-expert. 
In summary, our contributions in this work are two-fold: 1) proposing a DL framework by adding two channels comprising guiding signals to the selected nucleus and its neighboring nuclei 2) showing that the outputs from this framework can be useful in practice and for training deep networks.

\vspace{-.3cm}
\section{Methodology}
\vspace{-.3cm}
In the current work, we train NuClick model for different labeled datasets of nuclei. For each data set, based on the centroid of annotated nuclei, patches are extracted from larger images, and then two guiding channels are created to serve alongside RGB patches as the network input. The network's parameters are then optimized based on a weighted hybrid loss function. On the other hand, during the prediction phase, our framework accepts an image and its marked nuclei (clicked positions) from the user as inputs and generate the instance segmentation of the clicked nuclei in the output.
In the following, we describe each step in details.

\subsection{Model architecture and loss function}
\begin{figure}[t]
    \centering
    \includegraphics[width=\textwidth]{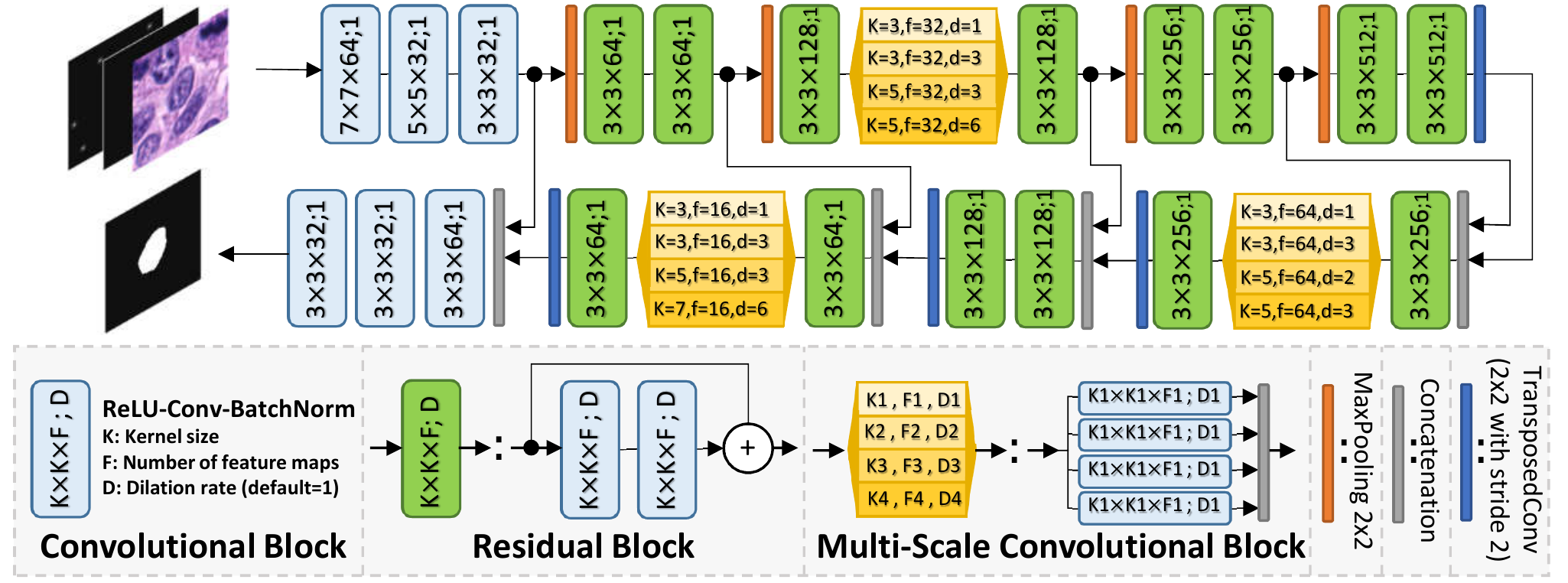}
    \caption{NuClick network architecture. Comprising convolutional, residual, and multi-scale blocks. Level transition is done using MaxPooling and TransposedConv layers.\vspace{-.5cm}}
    \label{fig:network}
\end{figure}

We have utilized an encoder-decoder architecture, inspired by U-net model{}, which reduces the size of feature maps in the encoding path while increasing their numbers, making abstract feature maps. The decoding path reverses this effect through several levels and turns those small and enriched feature maps into a single channel dense prediction. However, unlike the vanilla U-Net, the NuClick architecture incorporates residual and multi-scale convolutional blocks \cite{jahanifar2018segmentation} instead of normal convolutional layers in each level of encoding and decoding paths. An overview of the proposed NuClick architecture is depicted in Fig. \ref{fig:network}. 
Using residual blocks enables us to train the network with higher learning rates without being worried about gradient vanishing effect \cite{resnet}. Furthermore, multi-scale convolutional blocks allow the network to better capture the essence of image structures with different sizes and extract more relevant feature maps, hence boosting the network performance \cite{jahanifar2018segmentation}. 

For training the network, we proposed to use a hybrid weighted loss function, which is based on a soft variant of the Dice similarity coefficient and weighted binary cross-entropy (\ref{eq:loss}). 
The dice part of the loss controls pixel population imbalance problem during training as most of pixels belong to the background, and weighted binary cross entropy penalizes the loss if network wrongly segments the neighbouring nuclei. Our proposed hybrid loss is as follow:
\begin{equation}
\mathcal{L} = 1 - \frac{{\sum\limits_i {{p_i}{g_i} + \varepsilon } }}{{\sum\limits_i {{p_i} + \sum\limits_i {{g_i} + \varepsilon } } }} - \frac{1}{n}\sum\limits_{i = 1}^n {{w_i}({g_i}\log {p_i} + (1 - {g_i})\log (1 - {p_i}))} 
\label{eq:loss}
\end{equation}
where $\varepsilon$ is a small number, $n$ is the number of pixels in the image spatial domain, ${p_i}$, ${g_i}$, and ${w_i}$ are values of the prediction map, the ground-truths mask, and the weight map at pixel $i$, respectively. 

The pixel-wise weight map is generated based on the ground-truth, where regions of the neighboring nuclei have 10 times more weights than the desired nucleus (marked object). To better understand these maps, a simple image patch with the desired nuclei clicked (marked) in it, alongside its related ground-truth and weight map are illustrated in Fig. \ref{fig:maps}. We incorporated this weighting scheme for the loss function to put more emphasis on the neighboring nuclei and avoiding false segmentation of touching objects. In an alternative scenario, if we set the weights of the desired nuclei higher than other nuclei, the network may falsely get biased toward over-segmentation of neighboring nuclei.
\vspace{-.3cm}
\subsection{Guiding signal maps}

As a guiding signal and prior knowledge for the network, an extra channel is concatenated with the image in the network input which contains a 2D Gaussian distribution centered on the selected point (similar to \cite{koohababni2018nuclei} where the nucleus centroid was assumed as a Gaussian). We call this guiding signal the inclusion map, which refers to the nucleus we wanted to be included in the segmentation output. Adding the inclusion map helps the network in achieving a desirable segmentation of the selected nucleus as long as it is isolated. Based on our early experiments, when using only the inclusion map and there is a cluster of nuclei where the boundaries of each nucleus have overlapping or merged, the segmentation result may contain the touching nuclei.

\begin{figure}[t]
    \centering
    \includegraphics[width=\textwidth]{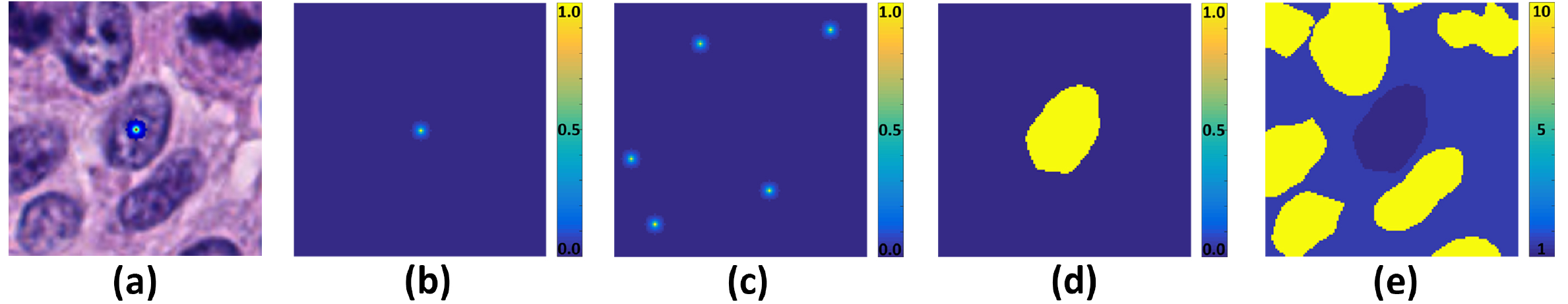}
    \caption{Guiding signal maps: (a)-(c) show inputs to the NuClick network which are image patch, inclusion map, and exclusion map, respectively, (d) depicts the desired network output (ground truth), and (e) illustrates pixel-wise weight map used in the loss function.\vspace{-.5cm}}
    \label{fig:maps}
\end{figure}

To avoid this phenomenon and to exclude the neighboring nuclei in the output prediction map, we introduce the exclusion map as the fifth channel to the input which can contain multiple 2D Gaussian distributions centered on the clicked positions of the neighboring nuclei (if annotator provides them). To get a better acquaintance with the inclusion and exclusion signal maps, they are visualized in Fig. \ref{fig:maps}(b)-(c) for a sample nuclei patch, also they can be seen at the input of the network in Fig. \ref{fig:network} as they are concatenated with the RGB image patch.

To this end,  the inclusion channel always provides a guiding signal for segmenting the desired nucleus and if other nuclei in the vicinity of the patch are selected the exclusion mask will also provide a signal; otherwise it is an all-zero channel. Please note that within the training phase the inclusion and exclusion maps are generated on-the-fly, based on the augmented (changed) ground-truth mask, and during the test phase they are directly constructed based on the user clicked positions.
\vspace{-.3cm}
\subsection{Training Procedure}

To optimize the network weights, an Adam optimizer with initial learning rate of 0.003 have been used. NuClick has been trained for 300 epoch and batch size of 128 on all datasets.
At each iteration, centroid positions of the desired nuclei are randomly jittered, and accordingly, the inclusion and exclusion maps are created on-the-fly.  This makes the network more robust against the variations in the input position provided by the annotator.   
\vspace{-.3cm}
\subsection{Testing Procedure}

In the test time, for each input image, the user clicks on the nuclei for annotation, or the centroids are loading from a file. Afterward,  for all available coordinates, patches of size 128$\times$128 are extracted from image and inclusion, and exclusion maps are created as mentioned in the previous section. The NuClick will predict a nucleus segmentation for each click (patch). Then that prediction map will be converted to a binary map by thresholding, and objects with areas smaller than 10 pixels would be removed (based on the size of the smallest object in the data set). The optimal threshold value, $T=0.4$, is selected by testing a set of candidate values and evaluating the resulted binary maps. Moreover, for removing extra objects except for the desired nucleus inside the binary map, the morphological reconstruction operator has been used which needs a marker and a mask. The marker has its all pixels equal to 0 except for a single pixel at the centroid location which is set to 1. On the other hand, the binary map plays the role of the mask in the morphological reconstruction. Having all patches predicted and processed, we can fill an empty canvas at the origin coordinates of each path with the processed nuclei masks to generate the final instance segmentation map of the input image.
\vspace{-.3cm}
\section{Experiments and Results}

\subsection{Dataset}

We have utilized two publicly available datasets in this work. Kumar dataset \cite{kumar2017dataset} contains 30 images of size 1000$\times$1000 which have been extracted from WSIs in The Tissue Genome Atlas (TCGA). This dataset covers seven tissue types and contains a total of 21623 nuclei instances segmented. From this dataset, 16 images are separated for training. The second dataset is  CMP17 dataset \cite{vu2019methods} which consists of 32 images of size between 500$\times$500 to 700$\times$700 and a total of 7570 nuclei instances. Similarly, 16 images in CPM17 are used to extract patches as the training set.
\vspace{-.3cm}
\subsection{Experimental results}

To show the generalizability of the NuClick across an unseen dataset, Table \ref{table:general} demonstrates quantitative results of NuClick when trained on a certain dataset (first column), and tested on another one (second column). Points for testing on the unseen dataset were provided from the centroids of objects in GT however they have been randomly jittered by 5 pixels to simulate manual annotation. We have used six evaluation metrics: Aggregated Jaccard Index (AJI), general Dice similarity coefficient, object-wise Dice (Dice\textsubscript{Obj}), Segmentation Quality (SQ), Detection Quality (DQ), and Panoptic Quality (PQ=SQ$\times$DQ) measures. The AJI metric \cite{kumar2017dataset} measures the quality of instance-wise predictions, DQ is equivalent to F1-score and only quantify the quality of detection, SQ reflects the average of intersection over union (IOU) for detected object, Dice for evaluates the similarity of overall nuclei segmentation against the GT, and Dice\textsubscript{obj} measures the Dice coefficient for individual segmented nucleus. Comprehensive information about these metrics can be found in \cite{grahamhover}.   

We also compare NuClick to two other approaches: U-Net, which is deep learning-based (supervised) model, and the watershed, which is an unsupervised method. For fair comparison in case of U-Net, the detection map (Gaussian centered on each nucleus centroid) of nuclei is concatenated to RGB channels at the training and testing phases. Moreover, the watershed has been applied to the U-Net prediction to have instance-wise outputs. In the unsupervised framework, the marker-controlled watershed algorithm is applied to the gradient map of the image using the centroid points as the markers. 

As reported in Table \ref{table:general}, the NuClick shows worse performance when it is trained on CMP and then test on Kumar, which is due to some hard cases (Cancerous colon) in Kumar dataset. Overall, NuClick performance, according to all metrics, is much better than the other two baselines that prove the high generalization capability of the NuClick. In an ideal situation,DQ metric for NuClick should be equal to 1, as it is representing the detection quality of the method and we have already provided the model with GT centroid defections. Nonetheless, DQ metric for NuClick is less than (yet very close to) 1. The reason is that NuClick does not consider some input points as valid nuclei, or the predicted map is eliminated during the thresholding and post-processing procedures. However, both detection and segmentation quality of NuClick is much higher than other reported methods in Table \ref{table:general}, which is also obvious in PQ metric. Quality of NuClick generated annotations is also evident from Fig. \ref{fig:motiv} which illustrates output masks for different clicked points in five image patches of different organs.
\begin{table}[]
\centering
\vspace{-.3cm}
\setlength{\tabcolsep}{2pt}
\begin{tabular}{c|cccccccc}
\textbf{Model}             & \textbf{Train} & \textbf{Test} & \textbf{AJI}    & \textbf{Dice\textsubscript{obj}}   & \textbf{Dice}   & \textbf{SQ}     & \textbf{DQ}     & \textbf{PQ}     \\ \hline
\multirow{2}{*}{NuClick}   & CPM            & Kumar         & \textbf{0.7940} & \textbf{0.7937} & \textbf{0.8886} & \textbf{0.8001} & \textbf{0.9819} & \textbf{0.7856} \\
                           & Kumar          & CPM           & \textbf{0.8278} & \textbf{0.8278} & \textbf{0.9088} & \textbf{0.8361} & \textbf{0.9981} & \textbf{0.8180} \\ \hline
\multirow{2}{*}{U-net+WS}     & CPM            & Kumar         & 0.7544          & 0.7601          & 0.8648          & 0.7823          & 0.9796          & 0.7663          \\
                           & Kumar          & CPM           & 0.7812          & 0.7844          & 0.8903          & 0.8074          & 0.9945          & 0.8029          \\ \hline
\multirow{2}{*}{Watershed} & -              & Kumar         & 0.1892          & 0.1660          & 0.4023          & 0.6936          & 0.3965          & 0.2805          \\
                           & -              & CPM           & 0.1501          & 0.1327          & 0.3467          & 0.7078          & 0.4243          & 0.3046         
\end{tabular}
\caption{Generalization of NuClick across CPM \cite{vu2019methods} and Kumar \cite{kumar2017dataset} datasets in comparison with other methods.\vspace{-.2cm}}
\label{table:general}
\end{table}
\vspace{-.3cm}
 Moreover, to validate the quality of annotations generated by NuClick another experiment has been designed: we first train NuClick on CPM (Kumar) data, and then used the trained NuClick to generate labels for Kumar (CPM) dataset. Afterwards, we train U-Net \cite{ronneberger2015u}, FCN8 \cite{long2015fully}, and Segnet \cite{badrinarayanan2017segnet} models on NuClick's annotations for Kumar and CPM dataset. Performances of these models are compared against those of same models trained on GT annotations.  Table \ref{table:vsGT} reports the results for this analysis. In this table, the title of each main column represents name of the dataset that we apply our model on. Each sub-columns of GT and NuClick\textsubscript{CPM/Kumar} indicate whether GT annotations or NuClick generated instances were utilized for training each model. Note that always GT annotations are used for model evaluation. 
 
 In Table \ref{table:vsGT}, for all networks, we observe relatively same results from outputs based on  GT and NuClick annotations. For instance, when testing on Kumar dataset, Dice and PQ values resulted from FCN8 model trained on NuClick\textsubscript{CPM}'s annotations are  0.01 and 0.003 (insignificantly) higher than the model trained on GT annotations, respectively. This might be due to more uniformity of the NuClick generated annotations, which eliminate the negative effect of inter-annotator variations present in GT annotations. This example and negligible differences in metrics values for two scenarios in all cases, prove that labels provided by NuClick are good enough to train deep networks. Note that All hyper-parameters and the order of feeding patches during training are the same for all experiments.
 
 
\begin{table}[]
\vspace{-.3cm}
\setlength{\tabcolsep}{5pt}
\begin{tabular}{ccc|cc|cccc}
       & \multicolumn{4}{c|}{Kumar}                            & \multicolumn{4}{c}{CPM}                                           \\ \cline{2-9} 
       & \multicolumn{2}{c}{GT} & \multicolumn{2}{c|}{NuClick\textsubscript{CPM}} & \multicolumn{2}{c}{GT}             & \multicolumn{2}{c}{NuClick\textsubscript{Kumar}} \\ \hline
Models & \textbf{Dice}       & \textbf{PQ}        & \textbf{Dice}          & \textbf{PQ}           & \textbf{Dice}  & \multicolumn{1}{c|}{\textbf{PQ}}    & \textbf{Dice}          & \textbf{PQ}           \\ \hline \hline
U-net  & 0.8243      & 0.5047      & 0.8196         & 0.5012        & 0.8535 & \multicolumn{1}{c|}{0.5878} & 0.8458         & 0.5798        \\
Segnet & 0.8465      & 0.5238     & 0.8368         & 0.5178        & 0.8716 & \multicolumn{1}{c|}{0.6268} & 0.8775         & 0.6281        \\
FCN8   & 0.7952      & 0.4484     & 0.8064         & 0.4512       & 0.8426 & \multicolumn{1}{c|}{0.5998} & 0.8294         & 0.5904       
\end{tabular}
\caption{Comparative experiments on CPM \cite{vu2019methods} and Kumar \cite{kumar2017dataset} test set with models trained using GT and NuClick's predicted masks. NuClick subscript indicates the dataset that used for its training.\vspace{-.2cm}}
\label{table:vsGT}
\end{table}

\vspace{-.7cm}
 \subsection{NuClick in Practice}
 \vspace{-.2cm}
 \subsubsection{LYON19 Challenge}
 LYON19 is a scientific challenge on lymphocyte detection in immuno-histochemistry (IHC) sample images.
 Challenge organizers released a dataset comprising 441 images of IHC stained specimens of breast, colon, and prostate.
The most challenging aspect of this task is that organizers did not release ground truth detection labels for the data set and instead asked the participants to use their data to develop a method.
To develop a well-performing supervised method, particularly deep learning based models, annotated data is required. Therefore,  NuClick was used to generate labeled data
We transformed the centroid detection problem into a nuclei instance segmentation task, 
where for each image in the released data set, we randomly sampled a 256$\times$256 patches to collect a subset of 441 training members. Then, a non-expert user reviewed all the patches and clicked on the positive lymphocytes based on his imperfect assumptions, which did not exceed 3 hours to be done completely. Image patches and their corresponding clicked positions are then fed into the NuClick framework to construct the instance segmentation map. After constructing a synthesized ground truth for each image, we developed instance segmentation models on the LYON19 task. Extracted centroids from the output instances of our model were able to rank \nth{1} in the \href{https://lyon19.grand-challenge.org/evaluation/results/}{LYON19 challenge leader-board} achieving F1-score of 0.7951. This state-of-the-art result proves the fidelity of the NuClick generated masks once again and shows that NuClick can be used reliably in generating data sets for such tasks.
 \vspace{-.3cm}
 \subsubsection{PanNuke}
 Another use case of NuClick is in extending dataset in our previous work, PanNuke \cite{gamper2019pannuke}, which demonstrates a  pipeline for creating large classification and segmentation labels for nuclei. 
 In this project, we need to provide instance segmentation,  according to the approximate centroids of nuclei. Since using off-the-shelf segmentation methods can lead to false-positive regions, we have used NuClick to provide labels just for desired locations.
\vspace{-.4cm}
\section{Conclusion}

We have proposed a simple and practical method for collecting nuclear annotation in histology images. We showed that one click from the user is enough to segment the nucleus, which is effortless and quick to collect a large number of annotations.  Moreover, we have shown that the labels generated bu NuClick are of high quality that can be used for training deep networks.

\bibliographystyle{unsrt}

\def\bibfont{\footnotesize}
\vspace{-.4cm}
\bibliography{MyCollection}

\begin{thebibliography}{10}

\bibitem{lu2018nuclear}
Cheng Lu, David Romo-Bucheli, Xiangxue Wang, Andrew Janowczyk, Shridar Ganesan,
  Hannah Gilmore, David Rimm, and Anant Madabhushi.
\newblock Nuclear shape and orientation features from h\&e images predict
  survival in early-stage estrogen receptor-positive breast cancers.
\newblock {\em Laboratory Investigation}, 98(11):1438, 2018.

\bibitem{kumar2017dataset}
Neeraj Kumar, Ruchika Verma, Sanuj Sharma, Surabhi Bhargava, Abhishek Vahadane,
  and Amit Sethi.
\newblock A dataset and a technique for generalized nuclear segmentation for
  computational pathology.
\newblock {\em IEEE transactions on medical imaging}, 36(7):1550--1560, 2017.

\bibitem{vu2019methods}
Quoc~Dang Vu, Simon Graham, Tahsin Kurc, Minh Nguyen~Nhat To, Muhammad Shaban,
  Talha Qaiser, Navid~Alemi Koohbanani, Syed~Ali Khurram, Jayashree
  Kalpathy-Cramer, Tianhao Zhao, et~al.
\newblock Methods for segmentation and classification of digital microscopy
  tissue images.
\newblock {\em Frontiers in bioengineering and biotechnology}, 7, 2019.

\bibitem{koohbanani2019nuclear}
Navid~Alemi Koohbanani, Mostafa Jahanifar, Ali Gooya, and Nasir Rajpoot.
\newblock Nuclear instance segmentation using a proposal-free spatially aware
  deep learning framework.
\newblock {\em arXiv preprint arXiv:1908.10356}, 2019.

\bibitem{maninis2018deep}
Kevis-Kokitsi Maninis, Sergi Caelles, Jordi Pont-Tuset, and Luc Van~Gool.
\newblock Deep extreme cut: From extreme points to object segmentation.
\newblock In {\em Proceedings of the IEEE Conference on Computer Vision and
  Pattern Recognition}, pages 616--625, 2018.

\bibitem{yang2006nuclei}
Xiaodong Yang, Houqiang Li, and Xiaobo Zhou.
\newblock Nuclei segmentation using marker-controlled watershed, tracking using
  mean-shift, and kalman filter in time-lapse microscopy.
\newblock {\em IEEE Transactions on Circuits and Systems I: Regular Papers},
  53(11):2405--2414, 2006.

\bibitem{jahanifar2018segmentation}
Mostafa Jahanifar, Neda Zamani~Tajeddin, Navid Alemi~Koohbanani, Ali Gooya, and
  Nasir Rajpoot.
\newblock Segmentation of skin lesions and their attributes using multi-scale
  convolutional neural networks and domain specific augmentations.
\newblock {\em arXiv preprint arXiv:1809.10243}, 2018.

\bibitem{koohababni2018nuclei}
Navid~Alemi Koohababni, Mostafa Jahanifar, Ali Gooya, and Nasir Rajpoot.
\newblock Nuclei detection using mixture density networks.
\newblock In {\em International Workshop on Machine Learning in Medical
  Imaging}, pages 241--248. Springer, 2018.

\bibitem{grahamhover}
Simon Graham, Quoc~Dang Vu, Shan E~Ahmed Raza, Ayesha Azam, Yee~Wah Tsang,
  Jin~Tae Kwak, and Nasir Rajpoot.
\newblock Hover-net: Simultaneous segmentation and classification of nuclei in
  multi-tissue histology images.

\bibitem{ronneberger2015u}
Olaf Ronneberger, Philipp Fischer, and Thomas Brox.
\newblock U-net: Convolutional networks for biomedical image segmentation.
\newblock In {\em International Conference on Medical image computing and
  computer-assisted intervention}, pages 234--241. Springer, 2015.

\bibitem{long2015fully}
Jonathan Long, Evan Shelhamer, and Trevor Darrell.
\newblock Fully convolutional networks for semantic segmentation.
\newblock In {\em Proceedings of the IEEE conference on computer vision and
  pattern recognition}, pages 3431--3440, 2015.

\bibitem{badrinarayanan2017segnet}
Vijay Badrinarayanan, Alex Kendall, and Roberto Cipolla.
\newblock Segnet: A deep convolutional encoder-decoder architecture for image
  segmentation.
\newblock {\em IEEE transactions on pattern analysis and machine intelligence},
  39(12):2481--2495, 2017.

\bibitem{gamper2019pannuke}
Jevgenij Gamper, Navid~Alemi Koohbanani, Ksenija Benet, Ali Khuram, and Nasir
  Rajpoot.
\newblock Pannuke: An open pan-cancer histology dataset for nuclei instance
  segmentation and classification.
\newblock In {\em European Congress on Digital Pathology}, pages 11--19.
  Springer, 2019.

\end{thebibliography}
\end{document}